# Low Cost, Educational Internal Combustion Engine Electronic Control Unit Hardware-in-the-Loop Test Systems

Sertac Karaman[a], Levent Guvenc[b]

[a]Department of Aeronautics and Astronautics, 77 Massachusetts Avenue Cambridge, MA 02139, USA
[b]Automated Driving Lab, Department of Mechanical and Aerospace Engineering, 201 W. 19th Avenue, Columbus, OH 43210, USA

## ABSTRACT

Different hardware platforms and their associated real time operating systems that can be used in an educational laboratory for illustrating engine electronic control unit hardware in the loop testing are presented and compared in this paper. A Matlab graphical user interface prepared for generating synthetic crank and camshaft angular position sensor signals to be fed to the engine electronic control unit during hardware-in-the-loop testing is introduced. This graphical user interface is used to generate faulty sensor signals to check the response of the engine electronic control unit during hardware-in-the-loop simulation. Examples of faulty signals that can be generated with the graphical user interface are illustrated.

---

## 1. Introduction

Automotive control and mechatronics is increasing its presence and importance every day in today's road vehicles. The road vehicle is becoming a totally mechatronic system loaded with a large array of sensors and actuators that are monitored and controlled by a large number of electronic control units in a vehicle wide network. Engine control was one of the first successful applications of automatic control in road vehicles. Consequently, complicated engine electronic control systems have been around for decades. Their complexity is steadily increasing due to increased levels of conflicting demands on allowed levels of undesired emissions, fuel economy and driving comfort. The increased complexity of engine control systems is made possible by advances in embedded control hardware and the use of advanced control methods.

Engine management and control systems exist as code in an engine Electronic Control Unit (ECU) in production vehicles. The ever increasing demand to reduce product development times also applies to the engine management and control system development for a new engine or for an engine being adapted to a new vehicle. Road testing and dynamometer testing are very useful and indispensable tools but the need for faster development times coupled with the desire to test conditions that may not easily be created otherwise have necessitated the development of Hardware-in-the-Loop (HiL) testing of engine ECUs. Examples of some early engine ECU HiL test systems can be found in the references Kimura and Maeda (1996), Isermann et al (1999), Schaffnit et al (1998) and Hanselmann (1996). A later survey of engine control module HiL systems is presented in Schuette and Ploeger (2007). More recent work on HiL simulation of engine ECUs can be found in many references like Mamala et al (2013), Bagalini and Violante (2016), Hartavi et al (2016), Xia et al (2018). An ECU HiL system provides a safe and fast lab based environment for testing advanced control algorithms like disturbance observer control (Guvenc and Srinivasan, 1995) or repetitive control (Demirel and Guvenc, 2010), neuro-dynamic programming (Boyali and Guvenc, 2010), parameter space robust control (Emirler et al, 2014; Guvenc et al, 2017)) or Model Predictive Control (Emekli and Aksun-Guvenc, 2016).

Full scale engine ECU HiL test systems used in the automotive industry are expensive devices and cannot easily be afforded by educational institutions. This paper presents several low cost educational solutions for illustrating engine ECU HiL that can be formed using off-the-shelf hardware that is available in most educational labs. This paper presents design and comparative analysis of five different hardware platforms, with different levels of HiL simulation capability. Different software development environments and their capabilities are presented for these hardware platforms. Capabilities are classified in terms of the different ECU signals that can be covered in the HiL simulations as well as the maximum sampling rate that can be achieved for a given hardware platform.

A current low cost solution to engine ECU HiL simulation is to use a so-called 'laboratory car' or lab car. A lab car is an electronic device that generates the signals that an engine ECU would normally receive from the engine sensors and other subsystems of the vehicle during operation. A typical electronic lab car is not a very expensive device but it suffers from the necessity of having to manually adjust important variables like engine speed and pressure and temperature sensor values by turning knobs during HiL testing. Engine calibration engineers use this platform with a calibration software for monitoring the internal signals of the ECU for presence of possible faults.

In this study, the first achievement was to digitally produce the signals that a lab car generates. At this stage, special emphasis was placed on the generation of the crank and cam shaft speed signals and development of necessary software for programmable engine speed capability. Six other sensory signals (mainly temperature and pressure values) were also simulated and the injection signals were captured in some of the hardware environments.

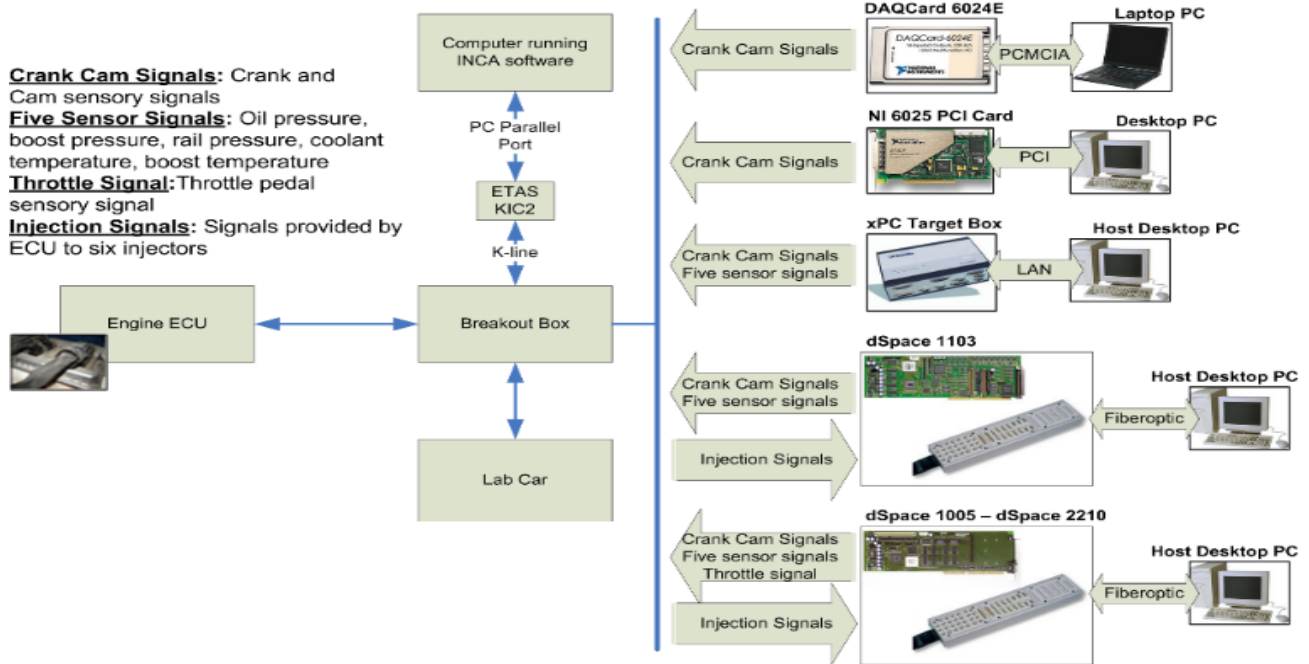

**Figure 1.** A general view of the hardware portion of the HiL setup showing all the hardware platforms used.

The second achievement was to develop a fault generator program for crank and cam shaft speed sensor signals. This program is able to handle simulation of broken, disturbed or amplified signals for individual crank or cam speed sensor teeth (pulses) as well as synchronization faults between crank and cam sensor signals. The faults can be prepared before the execution of the program and can be injected upon start of operation or they can be generated on the fly during the HiL simulation. Different ECU fault conditions can be diagnosed using a calibration software. The HiL simulator setups presented in this paper combine all these achievements and present students an environment to run different tests on the ECU, save and further analyze results automatically. Calibration for several crank and cam sensory signal faults are possible.

The organization of the rest of the paper is as follows. The HiL hardware setups used are presented in Section 2. Crank and camshaft signal generation and the injection of faults is treated in Section 3. The paper ends with conclusions in Section 4.

## 2. Hardware-in-the-loop test setups

The ECU used in the experimental tests was a commercial ECU for a six cylinder diesel engine with a common rail and turbocharger. The authors tested five different off-the-shelf hardware platforms (cards or systems used for signal generation and data acquisition) and their associated three different, readily available software platforms (real time operating system). The hardware platforms (see Figure 1) used were:

1. A laptop with a DAQCard 6024E PC card system,
2. A Personal Computer (PC) plus an NI6025E PCI card,
3. An xPC targetbox and a host PC,
4. A dSpace DS1103 card in an expansion box and a PC,
5. A dSpace compact simulator comprising of DS1005 processing and DS2210 signal generation cards and its connection box.

The software platforms used as the real time operating system all relied on the Simulink Real Time Workshop for automatic C code generation and were:

1. The Real Time Windows Target,

2. The xPC Target,
3. The dSpace RTI (Real Time Interface).

Along with the crank and camshaft position sensors, six other sensor signals were simulated. They were the throttle position, oil pressure, boost pressure, rail pressure, coolant temperature and boost temperature. These pressure and temperature sensors were modeled as look-up tables. Also, the six injection pulses sent by the ECU were captured by the HiL system.

In all of the five hardware setups above a calibration software was enabled for monitoring ECU internal signals and the presence of ECU faults. Lab Car hardware is also used for generation of the signals other than the ones generated by the simulation hardware interfaces. A general view of the HiL setup is illustrated in Figure 1.

The ECU is connected to a breakout box which allows easy access to individual ECU signals. The Lab Car connection to the ECU is used to supply signals that are not generated by the hardware platform used. If a signal is to be generated by one of the digital hardware platforms, then the Lab Car is bypassed with that signal and the breakout box connection is made through the desired hardware platform. Calibration software is run on a desktop PC and is connected to the ECU via the appropriate hardware module. This module connects the K-line bus of the ECU with the parallel port of the desktop PC. K-line bus connection is also made available on the breakout box. All five of the hardware platforms seen in Figure 1 were tested one at a time. Figure 1 also presents the signals that were generated and captured by each of the hardware platforms. The signals are partitioned into four groups; crank and cam signals, five sensory signals, throttle signal, injection signals. Each one of these groups is detailed in the same figure.

The capabilities of the hardware interfaces used in terms of the signals provided are constrained with the number of analog signals they are able to generate. For the platforms acquiring the injection signals, analog to digital conversion time is also important since it directly influences the quality of the captured signal. The hardware platforms are also constrained with their maximum analog output sampling period which directly affects the maximum engine rotational speed in rpm that can be reached during HiL simulation. CPU power of the hardware platforms is also an important issue since running the engine models in real-time requires considerable processing power depending on the complexity of the model. All these constraints are outlined in Table 1.

The DAQ6025E card has only 2 analog outputs and thus can only generate two of the signals. It was used for generation of crank and cam shaft sensory signals. The sampling time constrains the card to run up to an engine speed of 2,000 rpm in tests. The main advantage of the platform is that it is compact and hence portable. Since the software platform is the Real-Time Windows Target, the performance of the real-time simulation is directly related to the CPU power of the laptop PC used. It was observed in the tests that the system is slower than dedicated systems like xPC Target or dSpace systems regardless of laptop PC processing power. The authors relate this to the non-dedicated nature of the Real-Time Workshop Software platform.

**Table 1.** Hardware and software platforms

| Hardware Platform | Software Platform | I/O Sampling Rate | Maximum RPM reached | Number of analog inputs | Processing Power |
|---|---|---|---|---|---|
| DAQ Card 6025E | Real-time Windows Target | | | 2 | |
| NI PCI-6024 | Real-time Windows Target | 10 KHz | 2,500 rpm | 2 | Pentium 4 – 3 GHz |
| | xPC Target | | 2,500 rpm | | Pentium 4 – 3 GHz dedicated |
| xPC Target Box – Diamond MM 32 AT | xPC Target | ~166KHz (6us sampling time) | 2,000 rpm | 4 | Pentium 3 400 MHz dedicated |
| dSpace 1103 | dSpace RTI 1103 | 200 KHz | Over 5,400 rpm | 8* | Power PC 604 E @ 400 MHz dedicated |
| dSpace 1005 – dSpace 2210 | dSpace RTI 1005 dSpace RTI 2210 | ~50 KHz (20 us full-scale settling time) | Over 5,400 rpm | 12** | Power PC 750 @ 933 MHz dedicated |
| Hardware Platform | Software Platform | I/O Sampling Rate | Maximum RPM reached | Number of analog inputs | Processing Power |
| DAQ Card 6025E | Real-time Windows Target | | | 2 | |
| NI PCI-6024 | Real-time Windows Target | 10 KHz | 2,500 rpm | 2 | Pentium 4 – 3 GHz |
| | xPC Target | | 2,500 rpm | | Pentium 4 – 3 GHz dedicated |
| xPC Target Box – Diamond MM 32 AT | xPC Target | ~166KHz (6us sampling time) | 2,000 rpm | 4 | Pentium 3 400 MHz dedicated |
| dSpace 1103 | dSpace RTI 1103 | 200 KHz | Over 5,400 rpm | 8* | Power PC 604 E @ 400 MHz dedicated |
| dSpace 1005 – dSpace 2210 | dSpace RTI 1005 dSpace RTI 2210 | ~50 KHz (20 us full-scale settling time) | Over 5,400 rpm | 12** | Power PC 750 @ 933 MHz dedicated |

\* dSpace 1103 is also used for capturing injection signals, there are 20 analog inputs on the main processor board for this purpose.
\*\* dSpace 2210 also provides Angular Processing Unit (APU) for easy generation of crank and cam shaft sensor signals and acquisition of injection pulses. APU was used for crankshaft camshaft sensor signals and capturing injection pulses in the HiL simulations with dSpace 1005 instead of analog I/O.

The NI 6024 PCI card has two analog outputs. These analog outputs were used for generation of Crank and Cam signals in the tests. We also tested simulating two temperature or two pressure sensors at the same time. In these latter simulations, crank and cam signals were not generated by the NI 6024 PCI card. The software platform used was again the Real-Time Windows Target. The processing power of the hardware is again directly related to the CPU power of the PC used. The low performance in processing power can be observed again because of Real-Time Workshop being a non-dedicated software platform. It should also be noted that the NI 6024 PCI card is xPC Target compatible. Thus, the dedicated real-time operating system xPC Target can be used with this system provided that there exists another host computer for programming the desktop PC with NI 6024 PCI installed in it. Although this would increase the processing power, an improvement on maximum rpm was not possible as the hardware is constrained with a maximum 10 KHz sampling rate. More complicated models can still be executed when the xPC Target operating system is used.

The xPC Target Box is a PC-104 based computer system with Pentium III - 400 MHz CPU inside. This system is a dedicated system, i.e. no computation power of the host computer is used. xPC Target Box contains several different I/O cards allowing the unit to capture and generate analog signals, or capture encoder signals. It also has Controller Area Network (CAN) and RS-232 based serial communication peripherals. One of the most important advantages is being portable since the hardware is small-sized and of low-weight. For analog output, the most suitable I/O card was the Diamond MM-32 AT present on the xPC Target Box. The experiments showed that the xPC Target Box was quite slow in providing the necessary signals although the I/O card had a good sampling rate. The xPC target system with the NI 6024 PCI was even faster than the xPC Target Box system. The authors relate this to the ISA slots and PC-104 architecture of xPC Target box. The performance of the xPC Target system with NI 6024 PCI card can be further improved by using an I/O card with higher sampling rates. Although we have not experimented with such a platform, the NI PCI-6713 would be a nice solution, for example.

The dSpace 1103 system is a general-purpose simulation and controller hardware. The main processor is a Power PC 604E running at 400 MHz. I/O and features include A/D and D/A converters, digital I/O, serial interfaces (including RS232, RS422 and CAN) and incremental encoder interfaces. There is also a slave processor Texas Instruments TMS320F240 DSP running at 20 MHz which has several other I/O options. This slave DSP can be programmed for dedicated I/O operation while the main processor is doing the other complex computations. The dSpace 1103 unit contains eight analog outputs with high sampling rate. Using this device, we have experimented simulating crank and cam shaft sensory signals as well as oil pressure, rail pressure, boost pressure, coolant temperature and boost temperature signals. While all these signals were being HiL simulated, engine speeds up to 5,400 rpm were achieved. This is the maximum speed that is shown in the engine ECU monitoring software used. The crank and cam shaft sensory signals can be generated up to 8,000 rpm in simulations using this hardware platform. The analog inputs of dSpace 1103 were used for capturing the injections signals.

The final platform used in the experiments was the compact size simulator with dSpace 1005 processor card and dSpace 2210 I/O card. This is a special-purpose system for ECU and engine testing. It uses the Power PC 750 processor running at 933 MHz. The I/O card, dSpace 2210, contains a sensor and actuator interface providing analog and digital I/O, frequency and PWM measurement and generation as well as wheel speed sensor interface. It also provides serial interfaces (RS232, RS422 and CAN), and contains a slave DSP (Texas Instruments TMS320C31) for knock sensor simulation and wheel speed sensor simulation. The dSpace 1005 unit also contains an Angular Processing Unit (APU) that has several functions including generation of the crank and cam shaft sensor signals given the waveforms, and capturing the injection signals. For generation of crank and cam shaft sensor signals, the APU was used in this system instead of analog I/O channels. Analog output channels were used for simulation of other pressure and temperature sensor signals. Since the APU handles generation of the crank and cam shaft signals when the waveform is given, the CPU processing power can be used for running engine models or for other computational operations. This simulator unit is seen to be superior with very high sampling rates and higher dedicated processing power. This unit was able to run the diesel engine models with the ECU being in the loop.

## 3. Crank and cam shaft signal generation and injection of faults

The engine ECU used in this study controls a six cylinder engine. The crankshaft position sensor generates 60 peaks per revolution meaning a resolution of 6 degrees/peak. For indexing purposes, two of these peaks are empty. So, all together there are 58 pulse peaks and 2 null outputs in the crankshaft position sensor voltage as a function of crankshaft angle for one full revolution. The camshaft position sensor voltage output has 6 peaks corresponding to the six cylinders and one extra peak for indexing. Experimentally determined crankshaft and camshaft position sensor signals are shown in Figures 2 and 3, respectively.

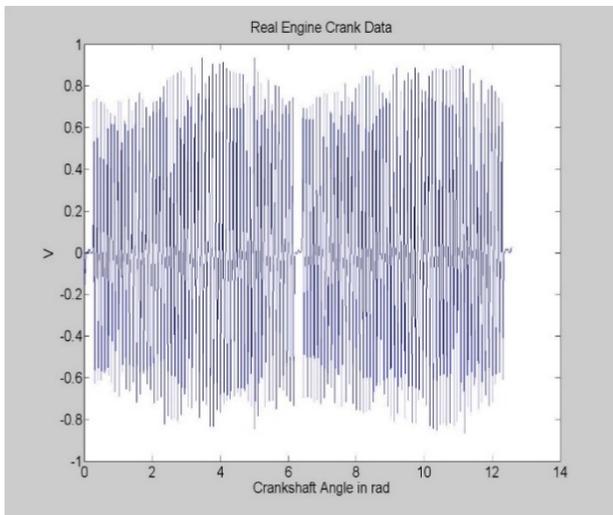

**Figure 2** Real engine crank shaft sensory data.

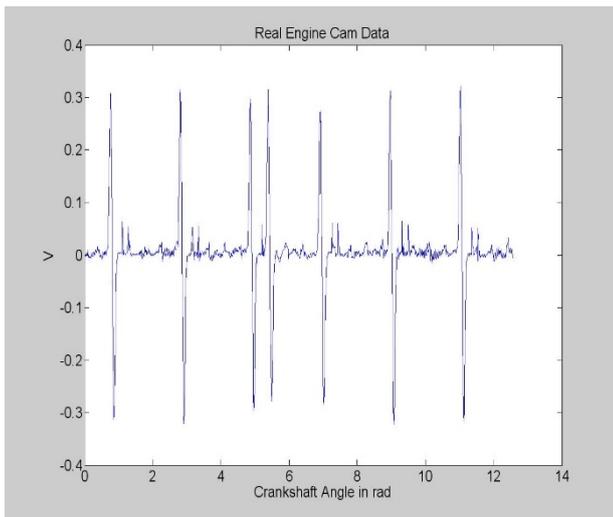

**Figure 3** Real engine cam shaft sensory data.

Generation of accurate and synchronized crankshaft and camshaft position sensor signals is vital for correct HiL simulation. Both signals are first created offline as a function of crankshaft angle. The crankshaft and camshaft signals as functions of crankshaft angle are shown in

Figures 4 and 5, respectively. The x axis in these two figures corresponds to two full crank revolutions. Note that the camshaft goes through one revolution for two revolutions of the crankshaft.

The crank and camshaft sensor waveforms are fixed (except for noise effects) as a function of crankshaft angle. These two waveforms are thus created and stored before the real time HiL simulation takes place. A separate program with a graphical user interface has been prepared to create these two waveforms offline, i.e. before the simulation. This program also allows the user to create crank and camshaft signals with a variety of faults. Possible faults which can be created by the program include missing peaks in the crank or camshaft sensor, changes in width or height of chosen parts of the signal (usually the peaks) and the addition of sensor noise. The program also allows user to partially or fully disturb a specific tooth (pulse) and generate synchronization faults between crankshaft and camshaft sensor data. It is also possible to inject these faults while the real time HiL simulation is running. The Graphical User Interface (GUI) of this program is shown in Figure 6. The graphical user interface of the sensor signal generation program is made up of a control panel on the upper left, a list box containing all the fault signals injected so far on the middle top, and a panel for easily opening, closing, compiling and downloading a simple Simulink model for basic tests. The main area of the graphical user interface shows the crankshaft and camshaft signals together. This can also be plotted in another figure window using the plot button on the interface. The control panel of the graphical user interface is shown in detail in Figure 7.

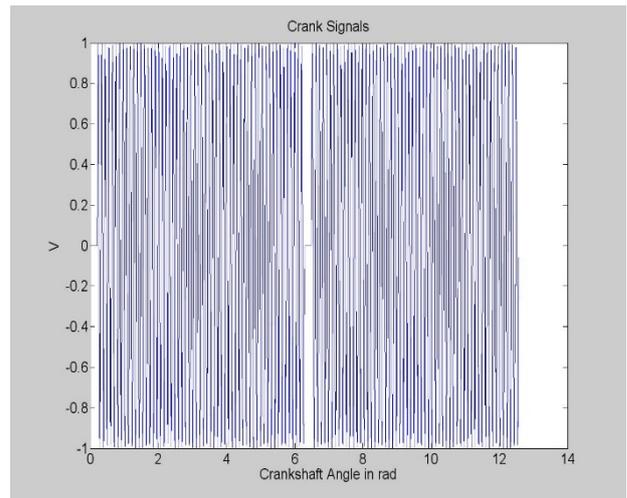

**Figure 4** Generated crankshaft position sensor signal.

Some examples of generated faulty signals are illustrated briefly in the following. Figure 8 shows broken teeth in the crank and cam angular position sensors in the form of missing pulses in the sensor outputs. The program illustrated in Figures 6 and 7 was used to generate signals with missing 27th crankshaft position sensor pulse and missing 2nd camshaft position sensor pulse. Figure 9 is a close-up view of Figure 8 that shows the defective crankshaft signal pulse more closely. The result in HiL simulation is a small error in the computed engine rotational speed.

Partially faulty crankshaft and camshaft sensor signal pulses are shown in the close-up view of Figure 10. The 28th crank and the 5th camshaft sensor pulses were distorted by the addition of small amplitude noise in Figure 10. The main sine wave form of the pulse can still be observed as the noise amplitude is small as compared to the sensor signal itself. The result in HiL simulation is a fault generated by the engine ECU.

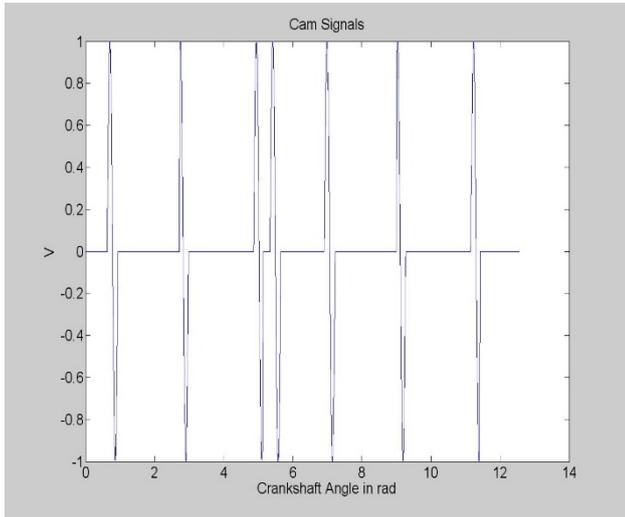

**Figure 5** Generated camshaft position sensor signal.

Partially faulty crankshaft and camshaft sensor signal pulses are shown in the close-up view of Figure 10. The 28th crank and the 5th camshaft sensor pulses were distorted by the addition of small amplitude noise in Figure 10. The main sine wave form of the pulse can still be observed as the noise amplitude is small as compared to the sensor signal itself. The result in HiL simulation is a fault generated by the engine ECU.

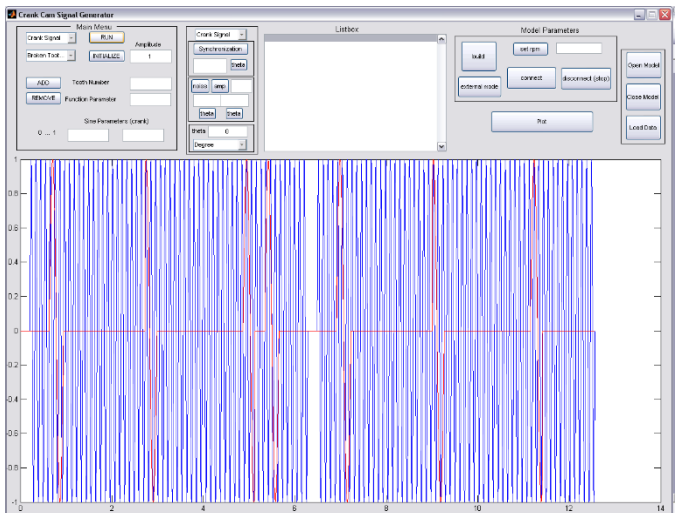

**Figure 6** Graphical user interface for generation of crankshaft and camshaft signal and injecting faults.

Faulty crankshaft and camshaft sensor signal pulses are shown in the close-up view of Figure 11. The 27th crank and the 2nd camshaft sensor pulses are distorted completely and are replaced by noise. The amplitude of the injected noise can be adjusted. The result in HiL simulation is a fault generated by the engine ECU.

## 4. Conclusions

Five different hardware platforms that can be used in an educational laboratory for illustrating engine ECU hardware in the loop testing were presented and compared in this paper. It is likely that one of these hardware platforms or a similar one will be present in the reader's lab, making the results presented here applicable. These hardware platforms work with different real time operating systems which were also presented. A Matlab graphical user interface prepared for generating synthetic crank and camshaft angular position sensor signals to be fed to the engine ECU during HiL testing was presented. This GUI can also be used to generate faulty sensor signals to check the response of the engine ECU during HiL simulation. Examples of some faulty signals that can be generated with the GUI were illustrated.

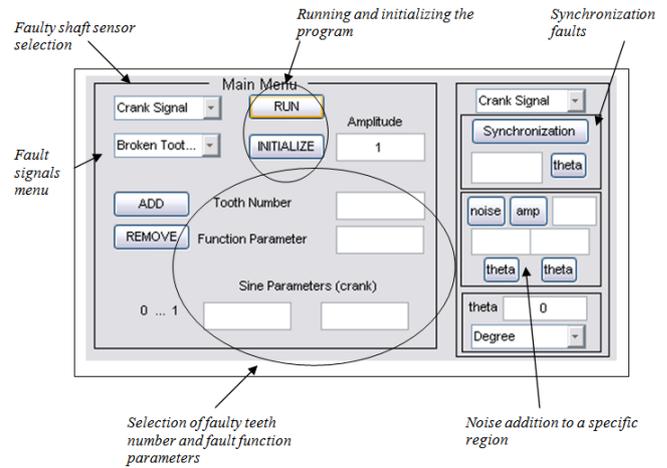

**Figure 7** Control panel of graphical user interface.

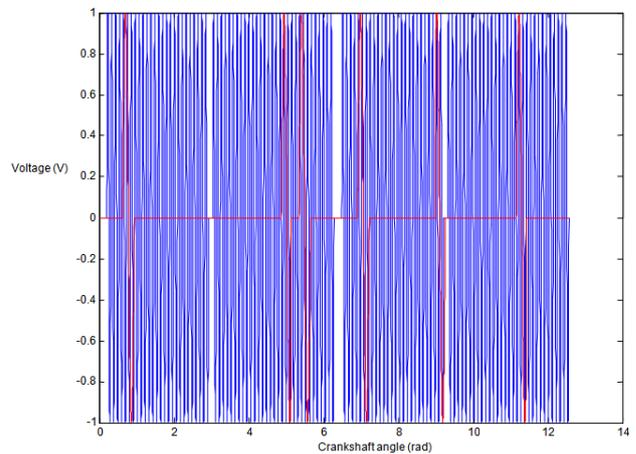

**Figure 8** Effect of broken sensor ring teeth for the crank and cam position sensor signals.

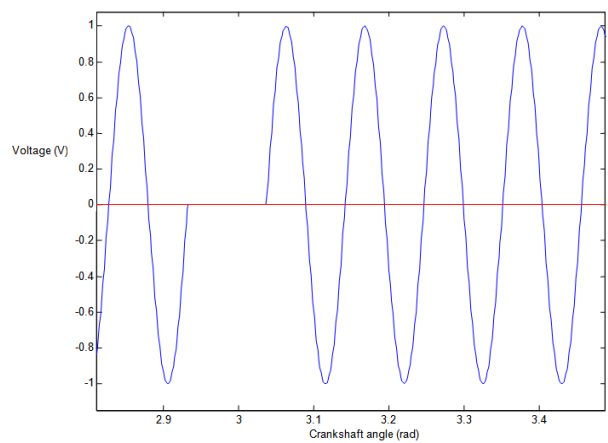

**Figure 9** Close-up view of Figure 8 showing broken sensor ring tooth for the crank position sensor signal.

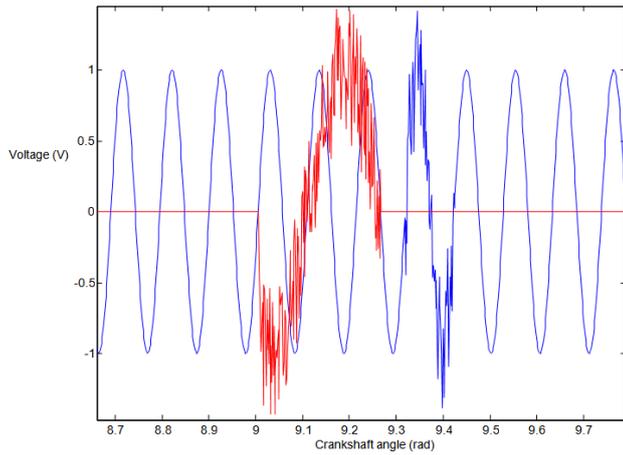

**Figure 10** Close-up view of generated partially faulty crankshaft and camshaft sensor signal pulses.

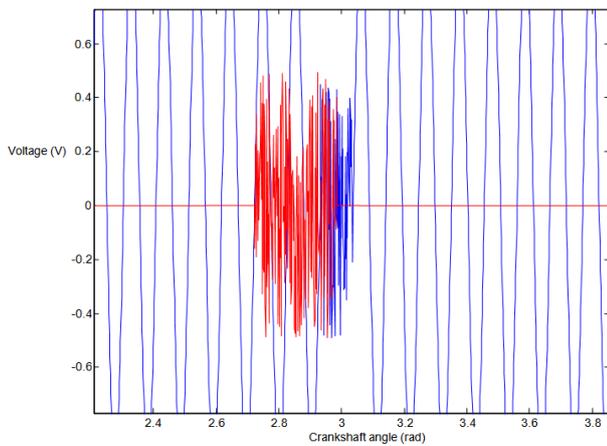

**Figure 11** Close-up view of generated faulty crankshaft and camshaft sensor signal pulses.